\documentclass[12pt,preprint]{aastex}
\usepackage{multirow}
 \newcommand{\beq}{\begin{equation}}
 \newcommand{\eeq}{\end{equation}}
 \newcommand{\beqn}{\begin{eqnarray}}
 \newcommand{\eeqn}{\end{eqnarray}}

\begin{document}
 
 \title{\bf{Possible Transit Timing Variations of  
the TrES-3 Planetary System}}
 
 \author{Ing-Guey Jiang$^1$, Li-Chin Yeh$^2$, 
Parijat Thakur$^3$, Yu-Ting Wu$^1$, Ping Chien$^1$, Yi-Ling Lin$^1$, 
Hong-Yu Chen$^1$, 
Juei-Hwa Hu$^1$, 
Zhao Sun$^{4,5}$, Jianghui Ji$^4$}
 
 \affil{
 {$^1$Department of Physics and Institute of Astronomy,}\\ 
 {National Tsing-Hua University, Hsin-Chu, Taiwan}\\
 {$^2$Department of Applied Mathematics,}\\
{National Hsinchu University of Education, Hsin-Chu, Taiwan}\\
{$^3$Applied Physics Unit,
Department of Basic Sciences \& Humanities, Guru Ghasidas
Central University, Bilaspur (C.G.) - 495 009, India}\\
{$^4$Purple Mountain Observatory, Chinese Academy of Sciences, 
Nanjing 210008, P.R. China}\\
{$^5$Graduate School of Chinese Academy of Sciences, 
Beijing 100049, P.R. China}
}
 
\begin{abstract}
Five newly observed transit light curves of the TrES-3
planetary system are presented. 
Together with other light curve data from literature,
23 transit light curves in total, which cover 
an overall timescale of 911 epochs, have been analyzed through a 
standard procedure.
From these observational data, the system's orbital parameters  
are determined and possible 
transit timing variations are investigated.
Given that a null transit-timing-variation produces 
a fit with reduced 
$\chi^2$=1.52, our results agree with previous work, that
 transit timing variations might not exist in these
data. However, a 1-frequency 
oscillating transit-timing-variation model, giving 
a fit with a reduced $\chi^2$=0.93, does possess a statistically higher
probability.
It is, thus, concluded that future observations
and dynamical simulations for this planetary system will
be very important.  
\end{abstract}
 
\keywords{planetary systems, 
stars: individual: TrES-3, techniques: photometric}
 
\section{Introduction}
 
The development of  research in extra-solar planetary systems 
has been successful for nearly two decades.  
In addition to the steadily increasing number of newly detected systems,
many new discoveries of planetary properties  also make  
this field extremely 
exciting. The Doppler-shift method makes many extra-solar
planetary systems known to us.  The methods of transit,
micro-lensing and direct imaging, also produce fruitful results. 
Particularly, 
more than 100 extra-solar planets (exoplanets) have been 
found to transit their host stars. Recently, based on the transit method,
the Kepler space telescope
has revealed many interesting results.
For example, one of the greatest achievement is the discovery 
of a system with six planets (Lissauer et al. 2011). 
These results truly demonstrate the
state-of-the-art power of the transit technique. 

Among these detected planetary systems, TrES-3 (O'Donovan et al. 2007)
attracts much attention due to its strong transit signal and 
short orbital period. 
For example,  Sozzetti et al. (2009) presented a work 
which combines the data from spectroscopic and photometric 
observations of TrES-3 to obtain the best models 
for the host star and planet. 
On the other hand, Fressin et al. (2010) used the Spitzer Space Telescope
to monitor TrES-3 during its secondary eclipse.
The most important constraint from these results is to 
show that the orbital eccentricity is very small, i.e. 
$|e\cos(\omega)| < 0.0056$, where $e$ is the orbital eccentricity
and $\omega$ is the longitude of the periastron.  
Furthermore, 
the 10.4 m Gran Telescopio Canarias 
(currently the world's largest optical telescope) has obtained 
extremely high-precision narrow-band transit data for the 
TrES-3 system (Colon et al. 2010).
There is almost no deviation for the data on the light curve, 
so this data could show  a strong constraint on orbital parameters of the 
exoplanet TrES-3b.


In order to study the perturbation from small unknown planets
and to constrain the overall orbital configuration in planetary systems,
transit timing variations (TTVs) have been seriously investigated  
in recent years (Agol et al. 2005, Holman et al. 2005).
Some of these works find that there is no TTV for the systems  
they studied. For example, Miller-Ricci et al. (2008) showed 
that there is no TTV signal above 80 s 
in the HD209458 system.   
Winn et al. (2009) also failed to confirm 
any TTV for the WASP-4 system. In contrast, 
Maciejewski et al. (2010) reported a successful case where
a periodical TTV was found and, which was likely due to an additional 15 
Earth-Mass planet located near the outer
1:2 mean-motion resonance in the WASP-3 system. 
Moreover, Holman et al. (2010)  confirmed TTVs in the double 
transiting planetary system Kepler-9.

For the TrES-3 system, there are also many studies
on TTV measurements. In addition to providing the best model
for the central star and planet,
Sozzetti et al. (2009) also found that 
the timing data gives a reduced $\chi^2$  of about 
5.87 with the assumption of no TTV. 
Authors concluded that there could be two possibilities: 
(1) the period is not a constant and, thus, it indicates some level of 
TTVs,or (2) the error bars might have been underestimated.
Gibson et al. (2009) showed that the reduced $\chi^2$ is 2.34 when the
period is set to be a constant. 
They also used the data to set the mass limit of the additional planet
by some three-body simulations.
Finally, Lee et al. (2011) obtained a reduced $\chi^2$ at about  2.48
for a linear fit with a constant period.

These results imply that further studies are necessary to 
address the significance of TTVs and explore a possible dynamical 
interpretation. 
Thus, in this paper, we plan to do further investigations
for the TrES-3 system
through a self-consistent procedure 
with both newly obtained 
observational data and  those from published papers.
It is the first time for the TrES-3 planetary system 
when 23 light curves, which cover a timescale of 911 epochs, 
are employed to determine orbital parameters and mid-transit times 
through a uniform procedure. The observational results are presented 
in \S 2, the analysis of light curves is shown in \S 3, the comparisons
with previous works 
are in \S 4, the possible TTV frequencies and models are discussed in \S 5 
and finally the conclusions are presented in \S 6.

\section{Observational Data}


\subsection{Observations and Data Reduction}
 
In this project, we monitor the TrES-3 system 
with the 0.8 m telescope at Tenagra Observatory
in Arizona, USA. In May and June 2010, five runs of 
transit observations  were successfully completed by our group.
For all these observations,the  R band was chosen and 
the exposure time was set to  100 s.
The field of view was 14.8 arcmin by 14.8 arcmin,
giving a scale of 0.87 arcsec/pixel.
The images had a typical full width at half maximum 
(FWHM) of 2.5 pixels.
A summary of the observations is presented in Table 1.
 
\vskip 0.1truein

\begin{table}[h]
\begin{center}
\begin{tabular}{llllll}
\hline
Run & UT Date & Interval (JD-2455000)& No. of Images  & Airmass Range 
& oot rms \\ \hline 
1 &2010 May 25 &  341.778 - 341.951  & 99 & 1.16-1.08 &  0.0019  \\
2 &2010 Jun 11 &  358.799 - 358.938  & 80 &  1.02-1.18 & 0.0020\\
3 &2010 Jun 15 &  362.716 - 362.856  & 80 &  1.17-1.02 & 0.0025\\
4 &2010 Jun 19 &  366.654 - 366.776  & 70 &  1.44-1.02 & 0.0018\\
5 &2010 Jun 28 &  375.779 - 375.917  & 77 & 1.01-1.29  & 0.0021 \\
\hline
\end{tabular}
\caption[Table 1]{The log of observations of this work.
For each run, the UT date, observational interval, 
number of exposures, airmass range 
and out-of-transit root-mean-square (oot rms) are listed.}
\end{center}
\end{table}

Using the standard IRAF software, 
images were first debiased and flat-fielded, 
and then the differential photometry was performed.
We searched for nearby bright stars with known identities
to be the candidates for comparison stars.
There were four candidates, but two of them were not 
suitable due to  the observed stellar flux being 
saturated or being variable. 
Thus, we finally used the star TYC3089-883-1 
and the star TYC3089-1137-1 as our comparison stars.
The flux of TrES-3 was divided by the flux of 
each comparison star, and also divided by the sum of 
the flux from both stars, to obtain three
light curves. The one with the smallest out-of-transit 
root-mean-square was used as the TrES-3 light curve for each 
transit event. The values of the out-of-transit 
root-mean-square (oot rms) of five transit light curves are listed in
Table 1, which quantify the quality of the light curves.  
The normalized relative flux and residuals as a function of the time 
of five transit light curves obtained from our
observations are shown in Figure 1.
 
\subsection{Other Observational Data from Literature} 

Eight light curves from  
Sozzetti et al.(2009), nine light curves from  
Gibson et al. (2009) and the transit data from Colon et al. (2010),
which employed the 10.4 m telescope, 
Gran Telescopio Canarias (GTC), 
are included in our analysis. 
{\it Please note that we do not simply use the mid-transit times
written in these papers, but analyze photometric data
with the same procedure and software 
to do parameter fitting in a consistent way.} 
This approach is a better procedure to study 
possible TTVs when different sources of data are considered.

\section{The Analysis of Light Curves}
 
The Transit Analysis Package (TAP) described by 
Gazak et al. (2011) was used for our 
light-curve analysis. In addition to Gazak et al. (2011),
Fulton et al. (2011) also presented the procedure
of using TAP in their study of a  HAT-P-13 system.
The Markov Chain Monte Carlo (MCMC)
technique and the model from Mandel \& Agol (2002) were employed to fit
the light curves in TAP. A  description
for other details of techniques about TAP can be found 
in the study by  Fulton et al. (2011).

Mandel \& Agol (2002) provided models of transit light curves 
derived from a simple two-body star-planet system. 
TAP can, thus, determine the best two-body orbital model 
from the observational data. In order to detect possible changes 
in orbital parameters, we analysed individual epochs separately.

Before using TAP,
the target's flux was normalized so that 
the out-of-transit flux values were close to unity. 
Moreover, as described in Eastman et al. (2010), 
TDB-based BJD was used for the time stamps in light curves here.

These light-curve data were then loaded into TAP to start MCMC chains.
For each light curve, five chains of length 300000 were computed. 
To start an MCMC chain in TAP, we needed to set the initial values of the 
following parameters: 
orbital period $P$,
orbital inclination $i$,
semi-major axis $a$ (in the unit of stellar radius $R_{\ast}$),
the planet's radius $R_{\rm p}$ (in the unit of stellar radius),
the mid-transit time $T_{\rm m }$, 
the linear limb darkening coefficient  $u_1$,
the quadratic limb darkening coefficient  $u_2$,
orbital eccentricity $e$ and the longitude of periastron $\omega$.
Once the above initial values were set, 
one could choose any one of the above to be:
(1) completely fixed (2) completely free to vary
or (3) varying following a Gaussian function, i.e. Gaussian prior,
during the MCMC chain in TAP.

We analyzed each light curve through TAP separately
and, thus, the orbital period was treated as a fixed parameter. 
The value of the orbital period in Sozzetti et al.(2009) was adopted. 
Because the orbital period is fixed, the semi-major axis 
shall not be completely free to vary. 
Thus, a Gaussian prior centred on the 
value 5.926
with $\sigma=0.056$ was set for $a/R_{\ast}$ during TAP runs, where
both the central value and the error bar were taken from 
Sozzetti et al.(2009). 

Moreover, $e$ and $\omega$ are simply fixed to be zero, and 
$i$ is treated in the same way as the semi-major axis.
We leave the mid-transit times $T_{\rm m}$ and $R_{\rm p}/R_{\ast}$ 
to be completely free during TAP runs and obtain
their best values for each light curve. They are the main parameters
we would like to obtain through light-curve data.
 
\vskip 0.1truein

\begin{table}[h]
\begin{center}
\begin{tabular}{lll} 
\hline
Parameter & Initial Value & During MCMC Chains \\
\hline  
$P$(days)& 1.30618581&     fixed \\
$i$(deg)  & 81.85         & a Gaussian prior with $\sigma=0.16 $        \\
$a$/$R_{\ast}$  & 5.926  & a Gaussian prior with $\sigma=0.056 $  \\
$R_{\rm p}$/$R_{\ast}$&0.1655&  free\\
$T_{\rm m }$ & set-by-eye &   free    \\
 $u_1$&Claret (2000,2004)&a Gaussian prior with $\sigma=0.05$\\
 $u_2$&Claret (2000,2004)&a Gaussian prior with $\sigma=0.05$\\ 
$e$              &   0.0    &  fixed  \\
 ${\omega}$ &   0.0   &  fixed    \\
\hline
\end{tabular}
\caption[Table 2]{
The parameter setting. The initial values 
of $P, i, a/R_{\ast}, R_{\rm p}/R_{\ast}$ are set as 
the values in 
Sozzetti et al.(2009)}
\end{center}
\end{table} 

For the limb darkening effect, 
a quadratic limb darkening law with
coefficients bi-linearly interpolated from
Claret (2000, 2004) to the values effective temperature 
$T_{\rm eff} =  5650.0$ K, stellar surface gravity 
log $g$ = 4.40 ${\rm cm/s^{2} }$,
metallicity [M/H] = -0.2 and  micro-turbulent velocity
 $V_{\rm t}$ = 2.0 km/s  is adopted, as in Sozzetti et al. (2009).
Thus, the limb darkening coefficients are treated as 
prior parameters. 
However, Southworth (2008) found that the difference between 
the best fitted 
limb darkening coefficients and those theoretical values interpolated from
Claret (2000, 2004) could be about  
0.1 or 0.2.
Thus, in order to take this possible difference into account, 
a Gaussian prior centred on the theoretical values 
with $\sigma=0.05$ is set for
limb darkening coefficients $u_1$ and $u_2$ during TAP runs, 
where
the value $\sigma=0.05$ is taken as  half of 0.1 and could 
make the Gaussian distribution's full width  
include possible differences. 
The details of parameter setting for TAP runs are listed in Table 2.

The limb darkening coefficients are dependent on 
the filters employed during observations.
Table 3 lists these theoretical
limb darkening coefficients $u_1$ and $u_2$ 
for all bands considered in this paper.
In any TAP run, 
these theoretical values are used as the  
initial $u_1$ and $u_2$, and also as the central values of
Gaussian priors.
However, the nine RISE light curves in Gibson et al. (2009)
were obtained through
a special instrument with a single non-standard wide-band filter
covering V and R bands.
The average values for filters V and R in Table 3 
were used as the theoretical 
limb darkening coefficients $u_1$ and $u_2$ 
when we ran TAP for these light curves.
Moreover, in Colon et al. (2010), 
two narrow near-infrared bands at 790.2 nm and 794.4 nm
are used to obtain high-precision transit data. 
There are 36 data points at 790.2 nm and   
35 data points at 794.4 nm, so in total 71 data points
are used for one transit event.
Since these two wavelengths are very close and
are at the centre of I band, 
the $u_1$ and $u_2$ for Filter I, shown in Table 3, were used
as the theoretical 
limb darkening coefficients.

\vskip 0.1truein

\begin{table}[h]
\begin{center}
\begin{tabular}{lll} 
\hline
filter & $u_1$  & $u_2$ \\
\hline  
 B &   0.6379 &   0.1792   \\
 V &    0.4378       &   0.2933        \\
 R &  0.3404     &    0.3190            \\
 I &  0.2576    &    0.3186      \\
 sloan u  &  0.8112    & 0.0554   \\
 sloan g  &  0.5535    &  0.2351  \\ 
 sloan r  &  0.3643    &  0.3178  \\
 sloan i  &   0.2777   &  0.3191 \\
 sloan z  &   0.2179   & 0.3162    \\
\hline
\end{tabular}
\caption[Table 3: The LD u1, u2]{
The theoretical limb darkening coefficients for the TrES-3 star.}
\end{center}
\end{table} 

All results  
derived through TAP are shown 
in Tables 4 and 5. The first TrES-3 transit in 
Sozzetti et al.(2009) is defined as epoch $E=0$,
and other transits' epochs are defined accordingly.
These two tables
list all parameter values with uncertainties following the order of epochs.  
Moreover, the observational light curves and 
best fitting models of our own data  
are presented in Figure 1, where 
the points are observational data and solid curves
are the best fitting models. The original data points in Figure 1
are available in a machine-readable form in the
electronic version of Table 6.

\vskip 0.1truein

\begin{table}[h]
\begin{center}
\begin{tabular}{cccccc}
\hline
Epoch &  Data Source &  $T_m$  & $i$(deg) & $a$/$R_{\ast}$ \\
\hline
0 & (a) & 4185.91111$^{+0.00021}_{-0.00021}$ &81.90 $^{+0.10}_{-0.11}$ & 
 5.906 $^{+0.044}_{-0.043}$ \\
10 & (a) & 4198.97359 $^{+0.00057}_{-0.00066}$ &
 81.79 $^{+0.14}_{-0.14}$ & 5.944 $^{+0.052}_{-0.052}$ \\
22 &(a) & 4214.64695 $^{+0.00032}_{-0.00036}$ &
81.81 $^{+0.12}_{-0.12}$ & 5.915 $^{+0.046}_{-0.045}$\\
23 &(a) & 4215.95288 $^{+0.00033}_{-0.00031}$ &
 81.77$^{+0.12}_{-0.13}$ & 5.937 $^{+0.047}_{-0.047}$ \\
267 &(b) &4534.66317 $^{+0.00019}_{-0.00019}$ &
 81.79 $^{+0.12}_{-0.12}$ & 5.952 $^{+0.044}_{-0.044}$\\
268 & (a) & 4535.96903 $^{+0.00039}_{-0.00037}$ &
81.83 $^{+0.12}_{-0.12}$ & 5.920 $^{+0.048}_{-0.047}$\\
281 &(a) & 4552.94962 $^{+0.00020}_{-0.00022}$ & 
 81.86 $^{+0.11}_{-0.11}$ & 5.939 $^{+0.043}_{-0.043}$\\
294 &(a) & 4569.92982 $^{+0.00039}_{-0.00040}$ & 
81.82$^{+0.12}_{-0.12}$ & 5.926 $^{+0.047}_{-0.047}$\\
313  &(a) & 4594.74682 $^{+0.00037}_{-0.00034}$ &
81.85 $^{+0.12}_{-0.13}$ & 5.943 $^{+0.047}_{-0.047}$\\
329 &(b) &4615.64621 $^{+0.00020}_{-0.00021}$ & 
 81.81 $^{+0.11}_{-0.11}$ & 5.915 $^{+0.044}_{-0.044}$\\
342 &(b) &4632.62690 $^{+0.00020}_{-0.00019}$ & 
81.81 $^{+0.11}_{-0.11}$ & 5.937 $^{+0.042}_{-0.042}$\\
355 &(b) &4649.60712 $^{+0.00019}_{-0.00017}$  & 
81.81 $^{+0.11}_{-0.11}$ & 5.931 $^{+0.042}_{-0.041}$\\
358 &(b) &4653.52661 $^{+0.00091}_{-0.00092}$ & 
81.89 $^{+0.14}_{-0.15}$ & 5.913 $^{+0.052}_{-0.052}$\\
365 &(b)& 4662.66984 $^{+0.00059}_{-0.00060}$  & 
 81.90 $^{+0.13}_{-0.14}$ & 5.915 $^{+0.051}_{-0.052}$\\
371 &(b) &4670.50709 $^{+0.00034}_{-0.00034}$ & 
81.86 $^{+0.11}_{-0.11}$ & 5.885 $^{+0.046}_{-0.045}$\\
374 &(b)& 4674.42521 $^{+0.00028}_{-0.00028}$ & 
 81.74 $^{+0.11}_{-0.12}$ & 5.965 $^{+0.047}_{-0.045}$\\
381 &(b)& 4683.56812 $^{+0.00042}_{-0.00041}$& 
81.85 $^{+0.12}_{-0.13}$ & 5.927 $^{+0.048}_{-0.048}$\\
665 &(c) &5054.52523 $^{+0.00018}_{-0.00017}$  &
81.83 $^{+0.10}_{-0.11}$ & 5.932 $^{+0.043}_{-0.043}$\\
885 &(d) &5341.8838 $^{+0.0011}_{-0.0010}$ & 
 81.81 $^{+0.15}_{-0.15}$ & 5.935 $^{+0.053}_{-0.053}$ \\
898 &(d) &5358.86606 $^{+0.00076}_{-0.00074}$ & 
81.75 $^{+0.14}_{-0.14}$
 & 5.957 $^{+0.053}_{-0.052}$\\
901 &(d) &5362.7847 $^{+0.0011}_{-0.00098}$ & 
 81.83 $^{+0.15}_{-0.15}$ & 5.937 $^{+0.053}_{-0.053}$\\
904 &(d) &5366.70215 $^{+0.00080}_{-0.00077}$ & 
81.95 $^{+0.14}_{-0.15}$ & 5.987 $^{+0.052}_{-0.052}$\\
911 &(d) &5375.84617 $^{+0.00089}_{-0.00090}$ & 
81.89 $^{+0.14}_{-0.15}$ & 5.912 $^{+0.052}_{-0.052}$\\
\hline
\end{tabular}
\caption[Table 4]{The results of light-curve analysis for 
$T_m$, $i$, and $a$/$R_{\ast}$.
Data sources: (a) Sozzetti et al.(2009),
(b) Gibson et al. (2009),
(c) Colon et al. (2010), and (d) this work.
To save space, the value of the mid-transit time $T_{\rm m}$ 
(in TDB-based BJD) is subtracted by 2450000. 
}
\end{center}
\end{table}

\vskip 0.1truein

\begin{table}[h]
\begin{center}
\begin{tabular}{cccc}
\hline
Epoch & $R_{\rm p}$/$R_{\ast}$ & $\mu_1$ &$\mu_2$ \\
\hline
0 &0.1656 $^{+0.0026}_{-0.0022}$ &0.219  $^{+0.049}_{-0.048}$ &
 0.317 $^{+0.050}_{-0.049}$ \\
10 & 0.1734 $^{+0.0075}_{-0.0068}$  & 0.642 $^{+0.049}_{-0.049}$ &
 0.183 $^{+0.049}_{-0.049}$\\
22 & 0.1708 $^{+0.0049}_{-0.0040}$ &0.440 $^{+0.049}_{-0.048}$ & 0.292 $^{+0.049}_{-0.049}$\\
23 & 0.1716 $^{+0.0058}_{-0.0049}$ &0.563 $^{+0.049}_{-0.049}$ & 0.242 $^{+0.048}_{-0.049}$\\
267 &0.1691 $^{+0.0041}_{-0.0034}$& 0.393 $^{+0.050}_{-0.050}$ &
0.312 $^{+0.050}_{-0.050}$\\
268 &  0.1639 $^{+0.0046}_{-0.0039}$ & 0.283 $^{+0.050}_{-0.049}$ & 0.322 $^{+0.049}_{-0.049}$\\
281 & 0.1634 $^{+0.0033}_{-0.0027}$ &0.350 $^{+0.049}_{-0.049}$
  & 0.308 $^{+0.050}_{-0.050}$\\
294 &0.1648$^{+0.0059}_{-0.0052}$ & 0.275 $^{+0.049}_{-0.049}$
& 0.317 $^{+0.050}_{-0.050}$ \\
313 & 0.1646 $^{+0.0053}_{-0.0046}$ & 0.273 $^{+0.050}_{-0.049}$
& 0.317 $^{+0.050}_{-0.050}$\\
329 &0.1674 $^{+0.0033}_{-0.0028}$& 0.400 $^{+0.049}_{-0.050}$ &
0.313 $^{+0.049}_{-0.049}$\\
342 & 0.1685 $^{+0.0036}_{-0.0031}$ & 0.390 $^{+0.048}_{-0.048}$
& 0.307 $^{+0.049}_{-0.049}$\\
355 &  0.1649 $^{+0.0030}_{-0.0026}$& 0.381 $^{+0.048}_{-0.049}$ & 
0.297 $^{+0.049}_{-0.049}$\\
358 & 0.1676 $^{+0.0084}_{-0.0074}$ & 0.385 $^{+0.049}_{-0.049}$
& 0.302 $^{+0.050}_{-0.049}$ \\
365&  0.1664 $^{+0.0071}_{-0.0061}$ &0.386 $^{+0.049}_{-0.049}$
& 0.303 $^{+0.049}_{-0.049}$\\
371 & 0.1619$^{+0.0032}_{-0.0029}$& 0.393 $^{+0.048}_{-0.049}$
   & 0.305 $^{+0.049}_{-0.048}$\\
374& 0.1616 $^{+0.0041}_{-0.0037}$ & 0.390 $^{+0.049}_{-0.049}$
        & 0.308 $^{+0.049}_{-0.049}$ \\
381& 0.1644 $^{+0.0053}_{-0.0044}$ & 0.387 $^{+0.050}_{-0.049}$
 & 0.305 $^{+0.049}_{-0.050}$\\
665 & 0.1655 $^{+0.0026}_{-0.0023}$  & 0.260 $^{+0.049}_{-0.049}$
& 0.320 $^{+0.049}_{-0.049}$ \\
885 & 0.1683 $^{+0.0097}_{-0.0086}$ & 0.345$^{+0.050}_{-0.050}$
& 0.323 $^{+0.050}_{-0.050}$\\
898 & 0.1627 $^{+0.0094}_{-0.0086}$ & 0.345 $^{+0.050}_{-0.050}$
& 0.324 $^{+0.050}_{-0.050}$\\
901& 0.160 $^{+0.011}_{-0.011}$& 0.341 $^{+0.050}_{-0.049}$
& 0.320 $^{+0.050}_{-0.050}$\\
904  & 0.1646$^{+0.0077}_{-0.0067}$& 0.334 $^{+0.050}_{-0.050}$
& 0.296 $^{+0.050}_{-0.050}$\\
911 &0.1543 $^{+0.0086}_{-0.0074}$  & 0.338 $^{+0.050}_{-0.049}$
& 0.316 $^{+0.050}_{-0.050}$\\
\hline
\end{tabular}
\caption[Table 5]{The results of light-curve analysis for 
parameters $R_{\rm p}$/$R_{\ast}$, $\mu_1$ and $\mu_2$.
}
\end{center}
\end{table}

\begin{table}[h]
\begin{center}
\begin{tabular}{ccc}
\hline
{\rm UT Date} & TDB-based BJD & Relative Flux \\
\hline
2010 May 25 & 2455341.781604507 &   0.996660 \\
      & 2455341.783375395  & 1.003077 \\
      & 2455341.785134680  & 0.998439 \\
\hline
2010 Jun 11 &  2455358.803032950 &  0.998805 \\
     & 2455358.804803775 &   1.002478 \\
&      2455358.806563056  &  0.997858 \\
\hline
2010 Jun 15 &    2455362.720387560  &  0.999725 \\
&      2455362.722158398 &  1.003418 \\
&      2455362.723940820 &  1.006198 \\
\hline
2010 Jun 19 &   2455366.657799268  &  0.998268 \\
&      2455366.659570101 &  0.997392 \\   
&  2455366.661329362 &  1.001116 \\
\hline
2010 Jun 28 &  2455375.782958203   & 0.997703 \\
&      2455375.784729044 &  0.999569 \\
&      2455375.786499864 &  0.999101 \\
\hline
\end{tabular}
\caption[Table 6]{The photometric light-curve data of this work. 
This table is 
available in its entirety in the on-line journal. 
A portion is shown here for guidance.}
\end{center}
\end{table}



For a photometric light curve, there are two
possible sources
of error. The uncorrelated Gaussian scatter
is called ``white noise'', and the time
correlated Gaussian scatter is ``red noise''.
TAP is designed to decompose and model the above
two sources of error with the technique
of  wavelet analysis.

The error bars of orbital parameters 
and mid-transit times shown in Table 4 
are the results of our TAP runs.
There are five chains in each of our TAP runs, 
and all of the chains are added together into the final results.
The 15.9, 50.0 and 84.1 percentile levels are recorded.
The 50.0 percentile, i.e., median level, is used as the best value, 
and the other two percentile levels give the error bar.

This error analysis
was tested and shown
to be successful in Gazak et al. (2011).
Moreover,
we found that when the deviations of transit-light-curve data 
are smaller, the resulting error bars become smaller.
Thus, the error bar does reflect the quality of data.
The MCMC procedure in TAP gives a reasonable estimation 
on the error related to the data itself.
Therefore, the error bars we obtained
here should  have been consistent with the
scattering and quality of the light curves, and provide
reliable error estimates.

\section{Comparison with Previous Studies}

Gibson et al. (2009) presented the results of transit timing residuals
in their Figure 3, where
both the data in Gibson et al. (2009) 
and Sozzetti et al. (2009) are included. Tables 3 and 4 in 
Gibson et al. (2009) provide the values of mid-transit times 
in HJD for all their nine 
transit light curves and also for those eight 
light curves from Sozzetti et al. (2009).
After converting these values from HJD to TDB-based BJD  
and adopting their values of uncertainties,
an ephemeris is calculated by minimizing $\chi^2$
through fitting a linear function: 
\beq
 T^C_{\rm m} (E)=  P E + T^C_{\rm m}(0), 
\eeq
where $P$ is the orbital period, $E$ is an epoch,  
$T^C_{\rm m} (E)$ is the calculated mid-transit
time at a given epoch $E$, and so $T^C_{\rm m}(0)$
is the value for $E=0$.
We found that 
$P=1.30618631\pm 0.00000016 $,
$T^C_{\rm m}(0)= 2454185.91111514\pm 0.00005033 $,
and the corresponding $\chi^2 = 34.26$.
Since the degree of freedom is 15, the reduced
$\chi^2 = 2.28$.  The $O-C$ diagram, which shows the differences 
between the observational  mid-transit time and 
the calculated mid-transit
time of a simple two-body star-planet system
(i.e. $T_{\rm m}-T^C_{\rm m}$)
as a function of epoch $E$, is 
shown in the top panel of Figure 2.

On the other hand, as we mentioned in \S 3, we
also have the mid-transit times directly derived from 
the light curves
in the study of  Sozzetti et al.(2009) and Gibson et al.(2009)
as shown in Table 4.
Similarly, by minimizing $\chi^2$
through fitting a linear function as Eq.(1), 
it was found that 
$P=1.30618691\pm 0.00000051$,
$T^C_{\rm m}(0)=2454185.91099199\pm  0.00015123 $,
and the corresponding
$\chi^2 = 19.69  $.
As the degree of freedom is 15, the reduced
$\chi^2 = 1.31$.
Thus, another $O-C$ diagram can be plotted as in the bottom panel
of Figure 2.

From both panels in Figure 2 and the resulting ephemeris,
we see that what we derived directly from 
light curves is correct. Our error bars are larger than
those in the previous work, so that the value of 
$\chi^2$ is smaller. Therefore, the error bars are unlikely 
to be underestimated in our results presented in this paper.

\section{Transit Timing Variations} 

\subsection{A New Ephemeris}
 
When 23 light curves mentioned in \S 2 are all considered, 
we would get a new 
ephemeris  by minimizing $\chi^2$
through fitting a linear function as Eq.(1).
We find that  
$P=1.30618619\pm 0.00000015 = P_l\pm 0.00000015 $,
where $P_l\equiv 1.30618619$, 
$T^C_{\rm m}(0)=
2454185.91116430\pm 0.00006123=T_l\pm 0.00006123$,
where $T_l\equiv 2454185.91116430$,  
and the corresponding 
$ \chi^2 = 31.95$.
Because the degree of freedom is 21, the reduced 
$\chi^2 = 1.52$. 
Using this new ephemeris,
the $O-C$ diagram is shown in Figure 3.

Therefore, for a straight line fit, i.e. a null-TTV model,
 our reduced $\chi^2$ 
with all considered 23 light curves is smaller than   
the value of 2.34 from Gibson et al. (2009).
This could be partially due to error bars here being slightly larger
than those in Gibson et al. (2009).
Therefore, our error bars are not underestimated,
and unlikely to lead to a result with false-positive TTVs.

\subsection{The Frequency Analysis and Possible Models}

In order to investigate whether there is any TTV,
Lomb's normalized periodogram (Press et al. 1992)
was used to search for possible variations
in the data.
Figure 4 shows the resulting spectral power as a function  of
frequencies. 
We defined the frequency with largest power as $f_1$,
i.e. $f_1\equiv 0.01055$, and tested the possible TTVs 
with frequency $f_1$ by 
minimizing $\chi^2$
through fitting a function as:
\beq
T_S (E) = P E+b+a\sin(2\pi f_1 E -\phi_1),
\eeq
where $T_S (E)$ is the predicted mid-transit time at a given epoch $E$,
$b, a, \phi_1$ are fitting parameters. 
We obtained that 
$P= 1.30618631\pm 0.00000031 $, 
$b= 2454185.91100290\pm  0.00012187 $,
$a=0.00036500 \pm  0.00009570$  
and $ \phi_1=3.97482109\pm 0.25632765 $.
The corresponding $\chi^2$= 17.59. Since the degree of freedom
is 19, the reduced $\chi^2$=0.93.
Using the above best fitted parameters for $T_S(E)$
and the new ephemeris $P_l$, $T_l$ for $T^C_{\rm m}(E)$,
the curve $T_S(E)-T^C_{\rm m}(E)$ as a function of epoch $E$ 
is plotted in the $O-C$ diagram, together with data 
points, as shown in Figure 5.

In order to test the models with multiple frequencies,
the frequency with the $i$-th highest power
is defined as $f_i$, where $i=2,...,5$.
Then, possible TTV models with $N$ frequencies can be tested by
minimizing $\chi^2$
through fitting a function as:
\beq
T_{S} (E) =  P E+b+
a \Pi_{i=1}^{i=N}\sin(2\pi f_i E-\phi_i),
\eeq
where $N=2,...,5$.  
These results of multiple frequencies, together with the results of 
null-TTV and one-frequency TTV models, are all 
summarized in Table 7. 
Similiarly, employing
the best fitted parameters for $T_S(E)$
and the new ephemeris $P_l$, $T_l$ for $T^C_{\rm m}(E)$,
the curves $T_S(E)-T^C_{\rm m}(E)$ as a function of epoch $E$
are shown in the $O-C$ diagram, together with data
points, as in Figure 6(A)-(D).

\begin{table}[h]
\begin{center}
\begin{tabular}{|c|c|c|c|}
\hline
Model & Fitted Parameters & Degree of Freedom & Reduced $\chi^2$  \\
\hline
null-TTV &  $P=1.30618619\pm 0.00000015$ &  21               &  1.52\\
   & $T^C_{\rm m}(0)=2454185.91116430\pm 0.00006123$&     &      \\
\hline
1-frequency & $P= 1.30618631\pm 0.00000031$   & 19    &  0.93 \\
        & $b= 2454185.91100290\pm  0.00012187 $   &        & \\
        & $a=0.00036500 \pm  0.00009570$   &        &  \\
   & $ \phi_1=3.97482109 \pm 0.25632765 $  &    &  \\
\hline
 2-frequency &$P=1.30618560\pm 0.00000045$ & 18  &  1.32 \\
         &$b= 2454185.91144686\pm 0.00016052$  &     &\\
         &$a=0.00053559\pm 0.00018953$  &    & \\
         &$\phi_1=2.46731638 \pm  0.28913521$ &     &\\
         &$\phi_2=5.58758211 \pm  0.31948870$ &    & \\
\hline
3-frequency &$P=1.30618619\pm  0.00000031$&  17  & 0.85  \\
        &$b=2454185.91111750\pm 0.00013402$  &      & \\
        &$a=0.00218541\pm 0.00062400$   &      & \\
        &$\phi_1= 0.49000000\pm 0.30458730$ &       &\\
        &$\phi_2=2.39100003\pm 0.13629963$ &       & \\
        &$\phi_3= 0.01779999\pm 0.07340797 $ &       &  \\
\hline
4-frequency & $P=1.30618572  \pm 0.00000036 $ & 16 & 0.86  \\
        & $b= 2454185.91134611 \pm 0.00018133 $    &    & \\
        &$a=0.00142777\pm 0.00059381 $    &    & \\
        &$\phi_1=  3.72499990\pm 0.31523481 $  &    &  \\
        &$\phi_2= 2.39299988\pm  0.42470514 $   &    &  \\
        &$\phi_3= 2.48769998 \pm 0.42087095 $    &    &   \\
        &$\phi_4= 0.27669999 \pm  0.28758257 $                        &    &   \\
\hline
5-frequency &$P=1.30618596\pm 0.00000040 $     &  15 & 0.64 \\
        &$b=2454185.91131390 \pm 0.00014342 $      &    &   \\
        &$a= -0.00219894\pm 0.00088575 $ &    &  \\
        &$\phi_1=3.68168663\pm 0.36549767 $    &   &   \\
        &$\phi_2=0.23282741 \pm 0.14959095 $    &    &   \\
        &$\phi_3=5.53840208 \pm 0.35394611 $                      &    &  \\
        &$\phi_4=2.04802298\pm 0.34912726 $      &   &   \\
        &$\phi_5=4.09944820\pm  0.33490377$     &      &  \\
\hline
\end{tabular}
\caption[Table 7]{The values of fitted parameters, 
degree of freedom, reduced $\chi^2$ of the null-TTV model and 
$i$-frequency models, where $i$=1,2,...,5.}
\end{center}
\end{table}

From the values of reduced $\chi^2$ presented in Table 7, it is clear that 
the model with one frequency, i.e. 1-frequency model, has the highest 
probability to produce the possible TTVs implied by the observational data.
In fact, 1-frequency, 2-frequency, 3-frequency and 
4-frequency models are
all better than the null-TTV model as their reduced $\chi^2$ are closer to 
unity. Among these,  
the curve of 3-frequency model almost goes through the error bars 
of all points. 
Due to the over-fitting,
the reduced $\chi^2$ of 5-frequency model
is very small.

\section{Concluding Remarks}
 
In this paper, five new transit light curves
and others from the literature, including 
one from Colon et al. (2010), eight from 
Sozzetti et al. (2009) and another nine from 
Gibson et al. (2009),
are employed to obtain the orbital parameters  
of the TrES-3 planetary system.
These 23 transit light curves, which cover an overall timescale
of 911 epochs, have been analyzed through a self-consistent standard 
procedure, so that the possible TTVs could be 
investigated carefully. 
We found that a null transit-timing-variation produced
a fit with a reduced 
$\chi^2$=1.52. 
This result agrees with the conclusions in previous work that
the transit timing variations might not exist in these
data. 

On the other hand, we also found that a 1-frequency 
transit-timing-variation model which gives 
a fit with a reduced $\chi^2$=0.93.
Thus, this interesting model is statistically more 
probable than the one with  null TTVs.
We conclude that the future high-precision observations
and further dynamical work for this planetary system 
will lead to fruitful scientific results. 
 
\section*{Acknowledgment}
The authors would like to thank the referee for 
good suggestions which greatly improved this paper. 
The authors are grateful to N.P. Gibson for kindly providing
the RISE light-curve data from Gibson et al. (2009), and 
A. Kong for the telescope time.  
This work was supported in part 
by the National Science Council, Taiwan, under 
NSC 98-2112-M-007-006-MY2 and NSC 100-2112-M-007-003-MY3.
PT gratefully acknowledges the support from
the Inter-University Centre for Astronomy and Astrophysics
(IUCAA), Pune, India, under the Associateship Programme.
J.J.H. acknowledges support by the National Natural Science
Foundation of China (Grants No. 11273068, 10973044, 10833001), 
the Natural Science Foundation of Jiangsu Province 
(Grant No. BK2009341), the
Foundation
of Minor Planets of Purple Mountain Observatory, and the
innovative and interdisciplinary program by CAS (Grant No. KJZD-EW-Z001).


\clearpage
\begin{figure}
\includegraphics[angle=0,scale=.80]{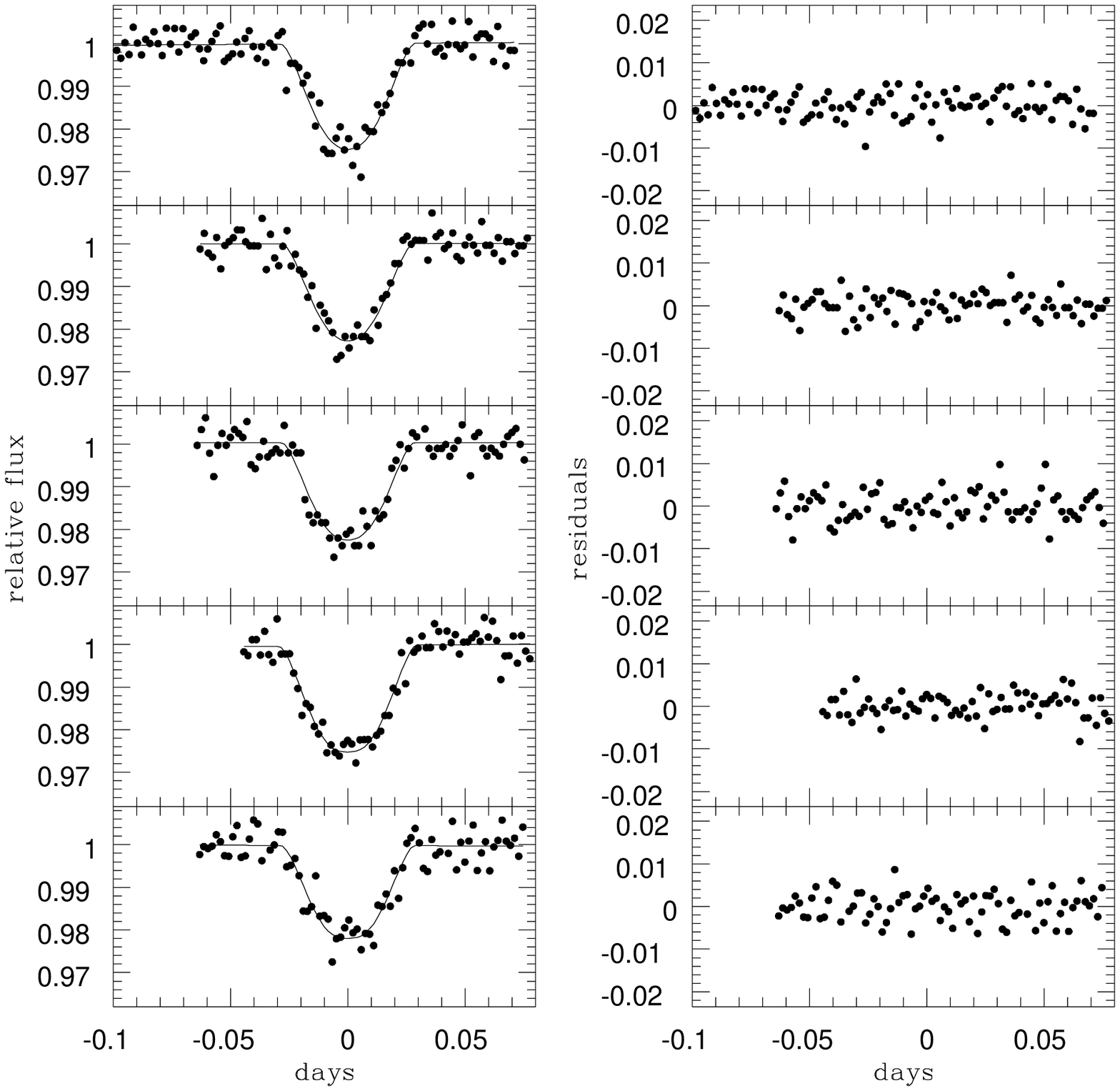}
\caption{(Left panels) The normalized relative flux as a function of 
the time (the offset from mid transit time and in TDB-based BJD) 
of five transit light curves of this work: points are the data and
curves are models. (Right panels) The corresponding residuals.
From top to bottom: our observational data of Run 1-5 listed
in Table 1.    
}
 \label {fig1}
 \end{figure}
 
 \clearpage
 \begin{figure}
\includegraphics[angle=0,scale=.80]{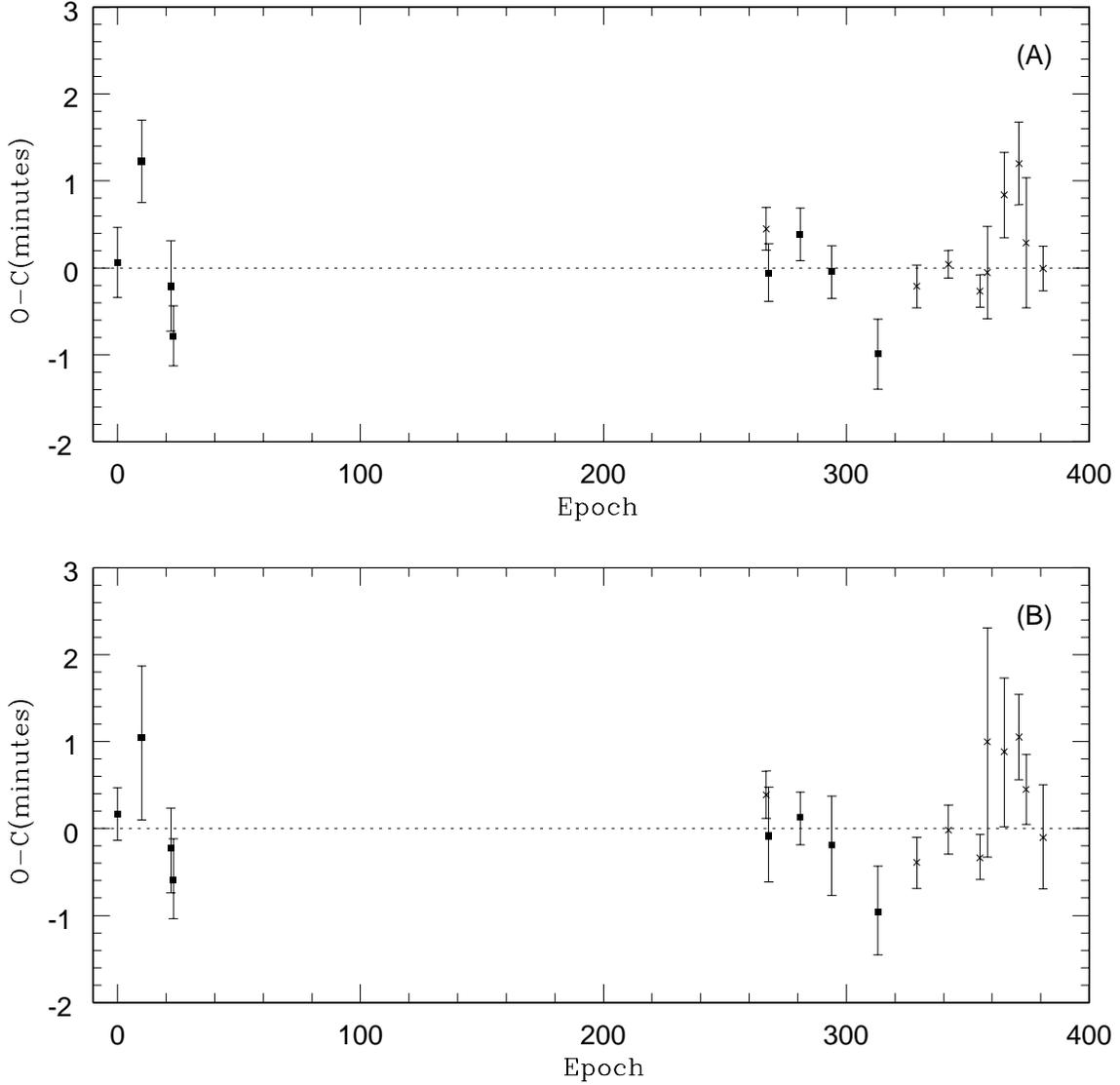}
\caption{The $O-C$ diagram. Panel (A) is for 
the results in which mid-transit times are taken from
Table 3 and Table 4 of Gibson et al. (2009) 
and converted to be in TDB-based BJD. 
Panel (B) is for the results of our analysis as those listed
in Table 4 but only those from Data Source (a) and (b) are 
included here. Squares are for the data from 
Sozzetti et al (2009), and crosses are for Gibson et al. (2009).
}
 \label {fig2}
 \end{figure}
 
 \clearpage
 \begin{figure}
\includegraphics[angle=0,scale=.80]{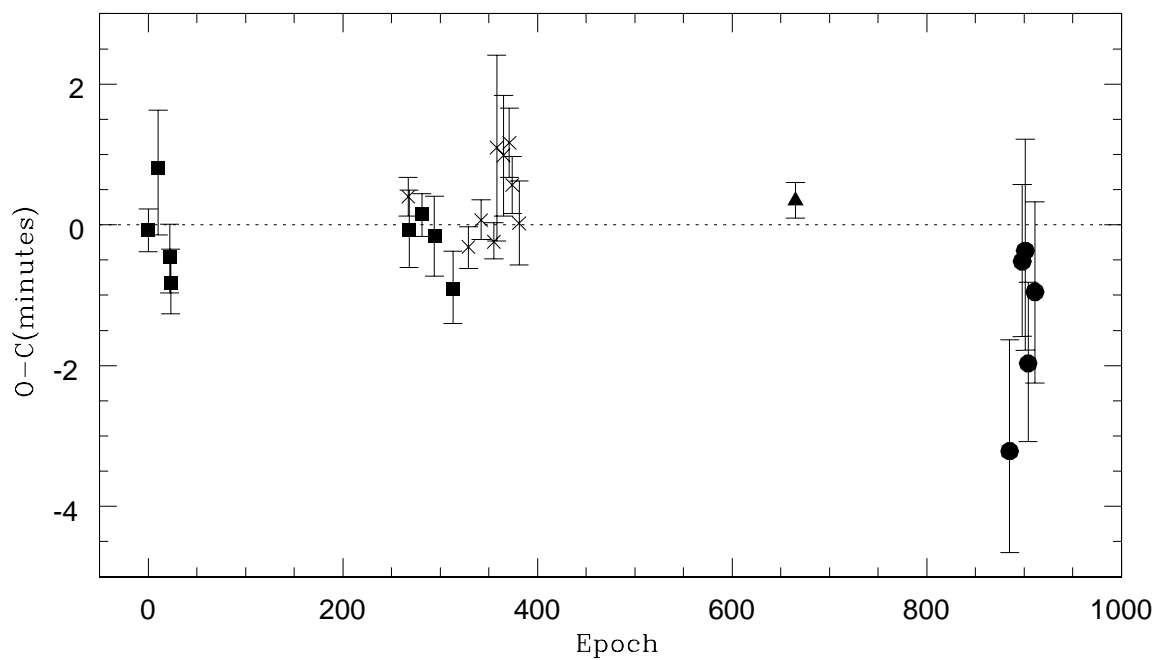}
\caption{The $O-C$ diagram for all results
shown in Table 4. 
The full circles are for this work, squares are for the data from 
Sozzetti et al (2009), the full triangle is for 10.4 meter GTC data,
and crosses are for the data from Gibson et al. (2009).
}
\label {fig3}
 \end{figure}
 
 \clearpage
 \begin{figure}
\includegraphics[angle=0,scale=.80]{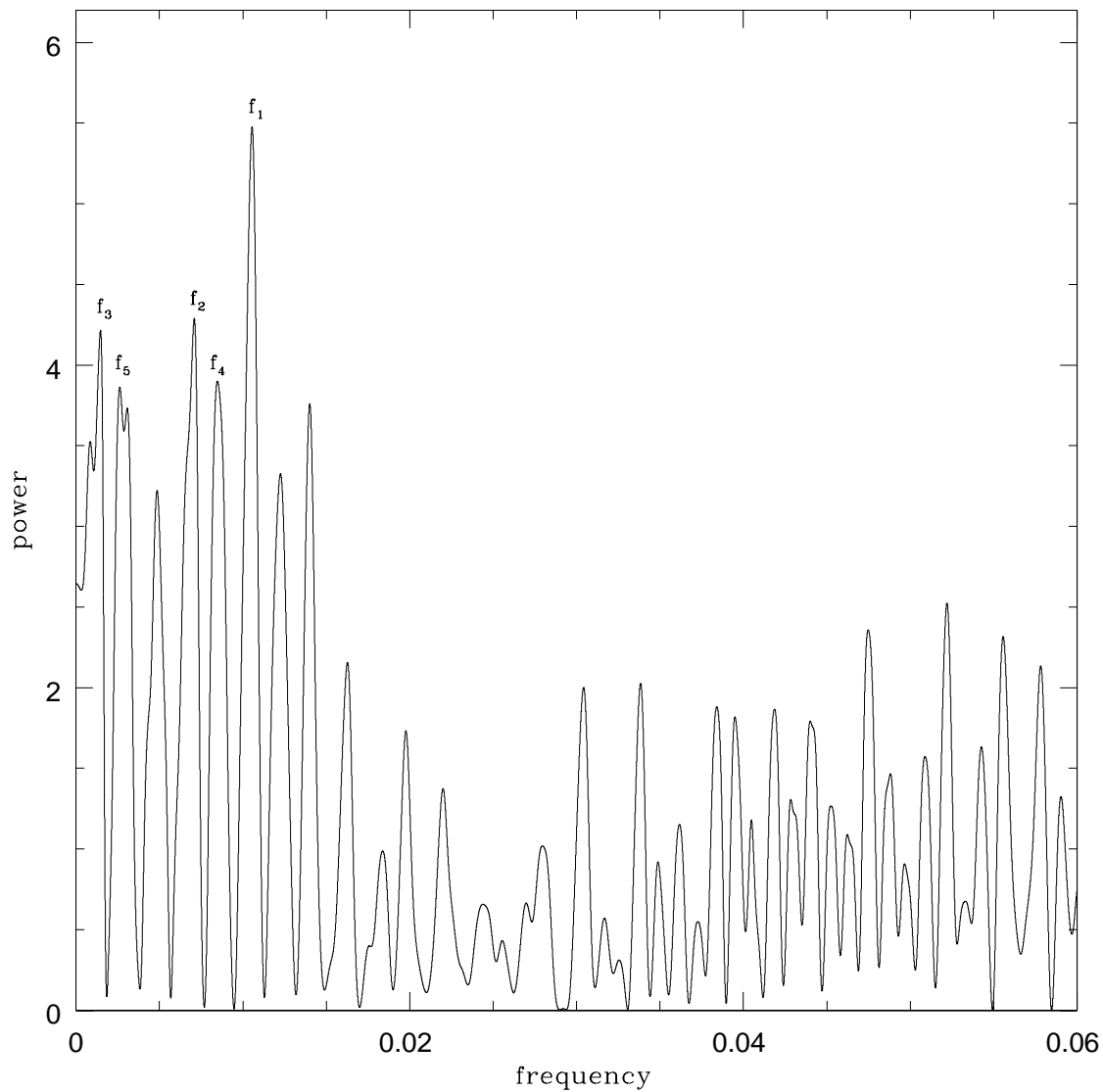}
\caption{The spectral power as a function  of
frequencies for the data points shown in Figure 3.
The top five frequencies are marked as 
$f_1, f_2, f_3, f_4$ and $f_5$.  
}
 \label {fig4}
 \end{figure}

\clearpage
\begin{figure}
\includegraphics[angle=0,scale=.80]{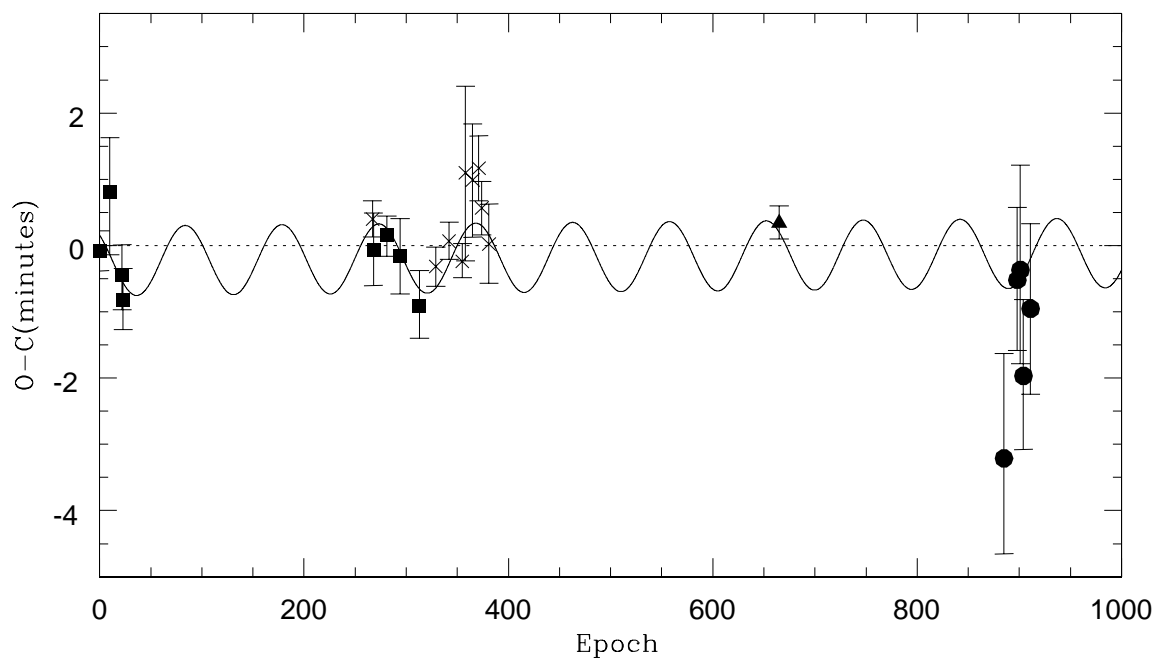}
\caption{The 1-frequency model and the $O-C$ diagram.
The curve is for the fitting function. 
The full circles are for this work, squares are for the data from 
Sozzetti et al (2009), the full triangle is for 10.4 meter GTC data,
and crosses are for the data from Gibson et al.(2009).
}
\label {fig5}
\end{figure}

\clearpage
\begin{figure}
\includegraphics[angle=0,scale=.80]{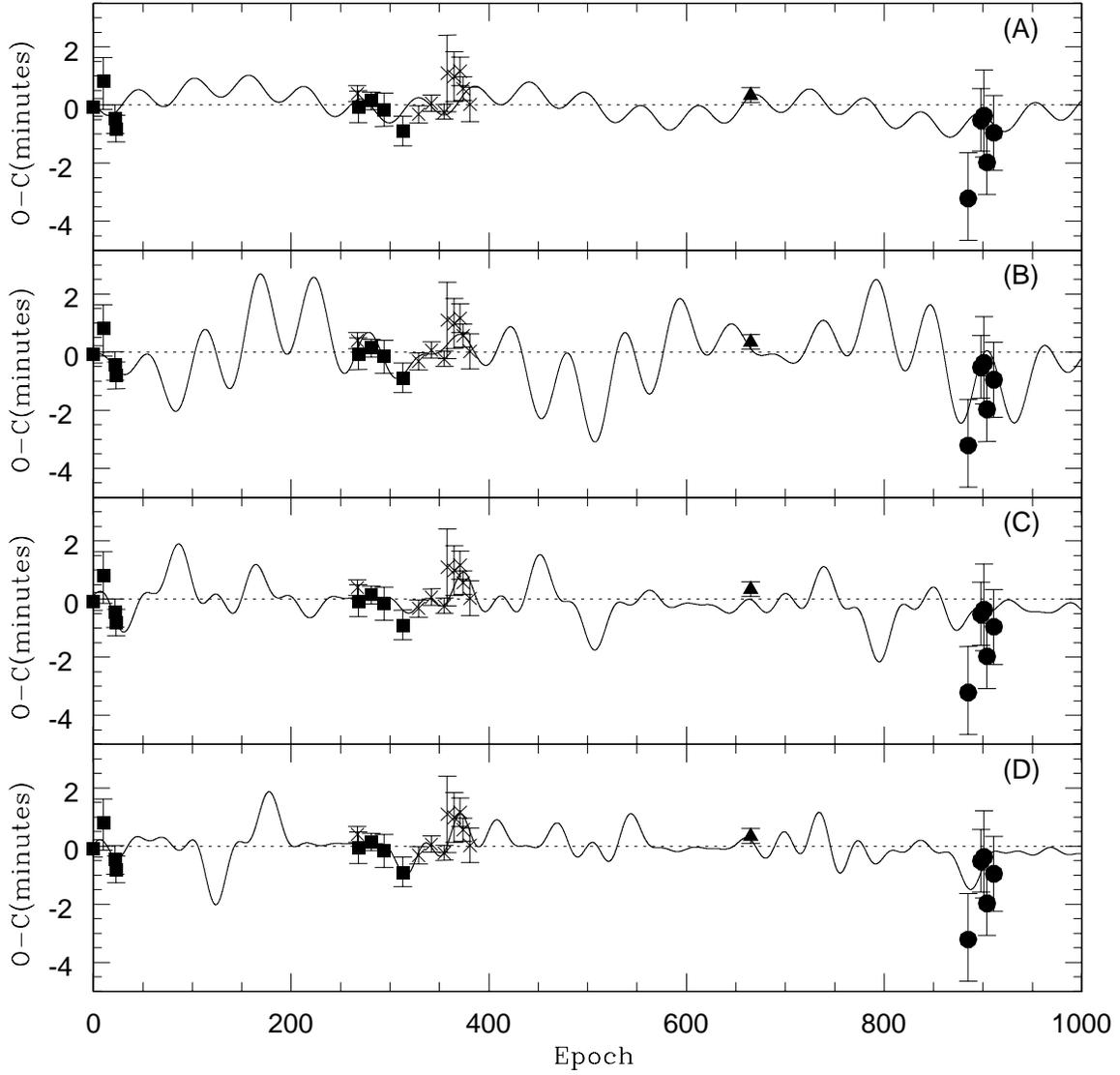}
\caption{The $i$-frequency model and the $O-C$ diagram, where 
$i$=2,3,4,5 (Panel (A)-(D)).
The curve is for the fitting function. 
The full circles are for this work, squares are for the data from 
Sozzetti et al (2009), the full triangle is for 10.4 meter GTC data,
and crosses are for the data from Gibson et al.(2009).
}
\label {fig6}
\end{figure}

\end{document}